\documentstyle[aps,epsfig]{revtex}

\date{today}

        \newcommand{\be}{\begin{equation}}
        \newcommand{\ee}{\end{equation}}
        \newcommand{\bea}{\begin{eqnarray}}
        \newcommand{\eea}{\end{eqnarray}}
        \newcommand{\ban}{\begin{eqnarray*}}
        \newcommand{\ean}{\end{eqnarray*}}
        \newcommand{\half}{\frac{1}{2}}

\begin{document}

\begin{center}
{\Large\bf The $O(N)$ Model at Finite Temperature: Renormalization
of the Gap Equations in Hartree and Large-$N$ Approximation}
\\[1cm]
Jonathan T.\ Lenaghan
\\ ~~ \\
{\it Physics Department, Yale University} \\
{\it New Haven, CT 06520, USA}
\\ ~~ \\ 
Dirk H.\ Rischke
\\ ~~ \\ 
{\it RIKEN-BNL Research Center, Physics Department} \\
{\it Brookhaven National Laboratory, Upton, New York 11973, USA}
\\ ~~ \\ ~~ \\
\end{center}

\begin{abstract}
The temperature dependence of the sigma meson and pion masses is studied
in the framework of the $O(N)$ model. The Cornwall--Jackiw--Tomboulis 
formalism is applied to derive gap equations for the masses
in the Hartree and large-$N$ approximations.
Renormalization of the gap equations is carried out within the cut-off and
counter-term renormalization schemes. 
A consistent renormalization of the gap equations within
the cut-off scheme is found to be possible only in the large-$N$ approximation
and for a finite value of the cut-off. On the other hand, 
the counter-term scheme allows for a consistent renormalization
of both the large-$N$ and Hartree approximations. 
In these approximations, the meson masses at a given nonzero temperature depend
in general on the choice of the cut-off or renormalization scale.
As an application, we also discuss the in-medium on-shell decay 
widths  for sigma mesons and pions at rest.
\end{abstract}

\section{Introduction}

Although chiral symmetry is manifest in the Lagrangian of
quantum chromodynamics (QCD) for
vanishing quark masses, quantum effects
break this symmetry spontaneously in the QCD vacuum.
At temperatures of order 150 MeV, however, lattice QCD results 
indicate that chiral symmetry is restored \cite{lattice}.
Such temperatures are expected to be reached
in ultrarelativistic heavy-ion collisions at CERN-SPS, BNL-RHIC, and 
CERN-LHC energies \cite{QM}.
The restoration of chiral symmetry may lead to observable 
consequences, for instance, in the dilepton mass spectrum \cite{dileptons}, 
or the formation of disoriented chiral condensates \cite{DCC}.

QCD with $N_f$ massless quark flavors has a $SU(N_f)_L \times
SU(N_f)_R$ symmetry. The order parameter for the chiral transition
is therefore $\chi^{ij} \equiv \langle \bar{q}^i_L q^j_R \rangle,
\, i,j = 1, \ldots, N_f$. 
For $N_f=2$, $SU(2)_L \times SU(2)_R$ is isomorphic to $O(4)$. Consequently,
the effective Lagrangian for the order parameter $\chi^{ij}$ falls in the same
universality class as the $O(4)$ model, with order parameter
$\phi^i, \, i = 1,\ldots, 4$.
Thus, if the chiral transition is second order in QCD, the dynamics
(and the critical exponents)
will be the same as in the $O(4)$ model, provided one is sufficiently close to
the transition temperature \cite{PisarskiWilczek}.
This motivates the study of the $O(4)$ model \cite{GellMann} as an
effective low-energy model for QCD.
The study which will be presented here is based on the $O(N)$ model for
arbitrary $N$. To make contact with QCD, however, we 
take $N=4$ in all numerical calculations.
 
At finite temperature, the naive perturbative expansion in powers of
the coupling constant breaks down, requiring resummation schemes to
obtain reliable results \cite{LindeKir,Dolan,Weinberg}. 
Typically, these schemes aim to include thermal 
fluctuations to all orders in the calculation of physical quantities.
Over the years, many ways to achieve this have been pursued
(some approaches are more rigorous, some are more or less {\em ad hoc}; 
Refs.\ \cite{GB,BraatenPis,Parwani,Bochkarev,Amelino,Roh,Ogure,Chiku,Berges}
constitute an incomplete list).
The present study of the $O(N)$ model focuses
on the Hartree approximation and its large-$N$ limit (the large-$N$ 
approximation).

The Hartree approximation is known from many-body theory
and generically represents the self-consistent resummation of tadpole diagrams 
\cite{FetterWal}. There is, however, no unique prescription to
perform this resummation.
This has led to the confusing situation that the same term 
``Hartree approximation'' has been used for resummation schemes which actually
differ in detail \cite{GB,Amelino,Roh}.
In this work, the so-called
Cornwall--Jackiw--Tomboulis (CJT) formalism is applied to
derive a Hartree approximation \cite{CJT}, and we shall use
the term ``Hartree approximation'' {\em exclusively\/} for 
the resummation of tadpoles within the CJT formalism.

The CJT formalism can be viewed as a prescription for computing the 
effective action of a given theory.
In general, the CJT formalism resums one-particle irreducible diagrams
to all orders. The stationarity conditions
for the effective action are nothing but Schwinger--Dyson
equations for the one- and two-point Green's functions of the theory.
What is usually \cite{Amelino,Petro} referred to as the Hartree approximation 
in the context of the CJT formalism is the special case where {\em only\/} 
one-particle irreducible {\em tadpole\/} diagrams are included in the 
resummation. (To be precise, the original work \cite{CJT} of Cornwall, 
Jackiw, and Tomboulis referred to this 
as the ``Hartree--{\em Fock\/} approximation''.)
In this case, the equations for the two-point Green's functions
simplify to self-consistency conditions, or ``gap'' equations, 
for the resummed masses of the quasi-particle excitations, i.e.,
in our case, the in-medium sigma and pion masses.

The $O(N)$ model has been previously studied using the 
CJT formalism by Amelino-Camelia \cite{Amelino}, and Petropoulos 
\cite{Petro}. The former work addressed renormalization using the
cut-off renormalization scheme, but did not present solutions of 
the gap equations. In the latter work, the gap equations were
numerically solved, but the issue of renormalization was not treated.
In this paper, we complete these investigations by discussing 
the renormalization of the gap equations in the cut-off as
well as the counter-term renormalization scheme and by presenting
the corresponding numerical solutions.

Renormalization of expressions obtained 
in self-consistent approximation schemes is non-trivial.
In perturbation theory, it is sufficient to perform
renormalization in the vacuum, $T=0$, order by order in the
coupling constant, as a finite temperature does not introduce
new ultraviolet singularities \cite{Kapusta}.
Consequently, the perturbative renormalization of the $O(N)$ model is
straightforward \cite{Peskin} and has been known for a long time
\cite{BWLee}. Self-consistent approximation schemes, however, 
in general resum only certain classes of diagrams.
This has the consequence that performing renormalization 
{\em after\/} resummation may require
renormalization constants that are no longer independent of the 
properties of the medium, see Ref.\ \cite{GB} and below. 
A resummation scheme which
circumvents this problem by renormalizing {\em prior\/} to
resummation is the so-called optimized perturbation theory
\cite{Chiku}. This approach will not be discussed here.

Our main results are the following. In the cut-off renormalization
scheme, we find that taking the cut-off, $\Lambda$, to infinity
the masses of the sigma meson and the pions become identical
\cite{Roh}, even in the phase where chiral symmetry is broken. 
This is clearly unphysical, as the pions are Goldstone bosons and 
thus much lighter than the sigma meson. Moreover, it is a well-known
fact \cite{Stevenson} that the $O(N)$ model becomes trivial in the
limit $\Lambda \rightarrow \infty$.
Consequently, renormalization of the $O(N)$ model within the cut-off
scheme can only be meaningfully studied for a finite value of $\Lambda$.
Even then, fixing $\Lambda$ to give the observed meson masses in the
vacuum, we find that in the absence of explicit chiral symmetry
breaking, the Hartree approximation
requires $\Lambda$ to vanish. The large-$N$ approximation does not
have this problem and allows for nonzero values of $\Lambda$.

In the counter-term scheme, renormalization can be performed within both the 
Hartree as well as the large-$N$ approximation. In the Hartree approximation,
the value of the renormalization scale, $\mu$, is uniquely fixed 
by the vacuum mass of the sigma meson in the absence
of explicit chiral symmetry breaking.
In the large-$N$ approximation,
this constraint does not exist, and $\mu$ can be chosen
arbitrarily.
As $\Lambda$ or $\mu$ are free parameters
in the large-$N$ approximation, at any given temperature the values of 
the meson masses depend on the choice of these parameters.
This is a consequence of the fact mentioned above that, 
for self-consistent resummation schemes, the renormalization constants
may depend on the temperature.

The outline of the paper is as follows.  
In Section II, we briefly discuss the effective potential 
within the CJT formalism \cite{CJT}. 
In Section III, this formalism is applied to derive the effective 
potential for the $O(N)$ model in the Hartree approximation.
Section IV is devoted to a discussion of the stationarity conditions 
for the effective potential, 
which lead to gap equations for the sigma and pion masses.
The renormalization of the gap equations is then performed 
in Section V within the cut-off and the counter-term schemes.
In Section VI we present numerical results.
Section VII concludes this paper with a summary of our results.
As an application we also compute the temperature dependence
of the in-medium decay widths to one-loop order for on-shell $\sigma$ 
and $\pi$ mesons at rest.

We use the imaginary-time formalism to compute quantities at
finite temperature. Our notation is 
\be 
\int_k \, f(k) \equiv T \sum_{n=-\infty}^{\infty} 
                       \int \frac{d^{3}k}{(2\pi)^{3}} \,
         f(2 \pi i n T,{\bf k}) \,\,\,\, , \,\,\,\,
\int_x \, f(x) \equiv \int^{1/T}_{0} d \tau \int d^{3}{\bf x} \,
f(\tau,{\bf x}) \,\, .
\ee 
We use units $\hbar=c=k_{B}=1$.  The metric tensor is $g^{\mu \nu}
= {\rm diag}(+,-,-,-)$.

\section{The Effective Potential in the Cornwall--Jackiw--Tomboulis 
formalism} 
\label{seceffpot}

The notion of an effective action is quite useful for studying
theories with spontaneously broken symmetries.  
For translationally invariant systems, the effective action 
becomes the effective potential.  At the classical level, 
the effective potential is given by the potential energy (density).
The vacuum (ground) state is given by the minimum of the potential 
energy.
For theories with spontaneously broken symmetry there may exist
infinitely many equivalent (degenerate) minima.
At the quantum level, there are additional terms in the 
effective potential, corresponding to 
quantum fluctuations. At finite temperature (and finite chemical potential),
the minimum of the effective potential corresponds to the thermodynamic 
pressure \cite{Rivers}.

The common way to compute the effective potential is via the loop expansion
\cite{Jackiw1}.
This approach, however, becomes problematic for theories with spontaneously
broken symmetries. In particular, the energy of quasi-particle excitations
with small 3-momenta becomes imaginary. The reason is that the requirement
of convexity for the effective potential is violated. A way to salvage the
loop expansion is to perform a Maxwell construction which restores the
convexity of the effective potential \cite{Rivers}. Another way to compute
the effective potential is via the CJT formalism
\cite{CJT}. As mentioned in the introduction, this method resums 
certain classes of diagrams
and has the advantage that the energy of the quasi-particle
excitations remains real for all values of 3-momentum.

Consider a scalar field theory with Lagrangian
\be \label{L}
{\cal L}(\phi) = \frac{1}{2} \, \partial_\mu \phi \, \partial^\mu \phi
- U(\phi) \,\, ,
\ee
for instance, $\phi^4$ theory where
\be
U(\phi) = \frac{1}{2} \, m^2 \phi^2 + \lambda \, \phi^4 \,\, .
\ee
The generating functional for Green's functions in the presence
of sources $J,\, K$ reads \cite{CJT}:
\be
{\cal Z}[J,K] = e^{{\cal W}[J,K]} =  \int {\cal D}\phi \, \exp
      \left\{I [\phi] +\phi \, J + \half \, \phi \, K \, \phi \right\} \,\, ,
\ee
where ${\cal W}[J,K]$ is the generating functional for
connected Green's functions, $I[\phi] = \int_x \, {\cal L}$ is
the classical action, and
\begin{mathletters}
\bea
\phi \, J & \equiv & \int_x \,\phi(x) \, J(x) \,\, , \\
\phi\, K \, \phi & \equiv & 
\int_{x,y} \,\phi(x) \, K(x,y)\,  \phi(y) \,\, .
\eea
\end{mathletters}
The expectation values for the one-point function, $\bar{\phi}(x)$,
and the connected two-point function, $G(x,y)$, in the presence of
sources are given by
\begin{mathletters}
\bea
\frac{\delta {\cal W}[J,K]}{\delta J(x)} & \equiv &\bar{\phi}(x) \,\, ,\\ 
\frac{\delta {\cal W}[J,K]}{\delta K(x,y)} & \equiv &
        \half \left[ G(x,y)+ \bar{\phi}(x) \, \bar{\phi}(y) \right] \,\, .
\eea
\end{mathletters}
One now eliminates $J$ and $K$ in favor of $\bar{\phi}$ and $G$ via a double
Legendre transformation to obtain the effective action
\be
\Gamma[\bar{\phi},G] = {\cal W}[J,K] - \bar{\phi} \, J - 
        \half \, \bar{\phi}\, K\, \bar{\phi} -  \half\, G\, K \,\, ,
\ee
where $G\, K \equiv \int_{x,y} \, G(x,y)\, K(y,x)$. Thus, 
\begin{mathletters}
\bea
\frac{\delta \Gamma[\bar{\phi},G]}{\delta \bar{\phi}(x)} & = &
        J(x) - \int_y \, K(x,y)\, \phi(y) \,\, ,\\   
\frac{\delta \Gamma[\bar{\phi},G]}{\delta G(x,y)} & = & -\half\,  K(x,y) \,\, .
\eea
\end{mathletters}
For vanishing sources, we find the stationarity conditions which determine
the expectation value of the field $\varphi(x)$ and
the propagator ${\cal G}(x,y)$ in the absence of sources:
\begin{mathletters}
\bea \label{phi}
\left. \frac{\delta \Gamma[\bar{\phi},G]}{\delta \bar{\phi}(x)} \, 
\right|_{\bar{\phi}=\varphi,\, G= {\cal G}} &  = & 0 \,\, , \\
\left. \frac{\delta \Gamma[\bar{\phi},G]}{\delta G(x,y)} 
\right|_{\bar{\phi}=\varphi,\, G= {\cal G}} &  = & 0\,\, . \label{G}
\eea
\end{mathletters}
Equation (\ref{G}) corresponds to
a Schwinger--Dyson equation for the full (dressed) propagator.
It was shown in \cite{CJT} that the effective action $\Gamma[\bar{\phi},G]$ 
is given by
\be 
\Gamma[\bar{\phi},G] = I[\bar{\phi}] - \half \,
        {\rm Tr}\left(\ln \,G^{-1}\right) - \half\, 
        {\rm Tr}\left(D^{-1}\, G-1\right) + \Gamma_2[\bar{\phi},G]\,\, . 
\ee
Here, $D^{-1}$ is the inverse of the tree-level propagator,
\be
D^{-1} (x,y;\bar{\phi}) \equiv - \left. \frac{\delta^2 I[\phi]}{\delta 
\phi(x) \,
\delta \phi(y) } \right|_{\phi = \bar{\phi}} \,\, ,
\ee
and $\Gamma_2[\bar{\phi},G]$ is the sum of all
two-particle irreducible diagrams where all lines represent full 
propagators $G$.

For constant fields $\bar{\phi}(x) = \bar{\phi}$, homogeneous systems,
and for a Lagrangian of the type given by eq.\ (\ref{L}),
the effective potential $V$ 
is given by $V = -T\Gamma/\Omega$, where $\Omega$ is the
3-volume of the system, i.e.,
\be \label{14}
V[\bar{\phi},G] = U(\bar{\phi}) +\half \int_k \, 
\ln G^{-1}(k) + \half\,\int_k\, \left[D^{-1}(k;\bar{\phi})
\, G(k) - 1 \right] + V_2[\bar{\phi},G] \,\, ,
\ee
with 
\be 
D^{-1}(k;\bar{\phi}) = -k^2 + U''(\bar{\phi})
\ee
and $V_2[\bar{\phi},G] \equiv - T \,\Gamma_2[\bar{\phi},G]/\Omega$.
The stationarity conditions are given by 
\begin{mathletters}
\bea \label{stationphi}
\left. \frac{\delta V[\bar{\phi},G]}{\delta \bar{\phi}} \, 
\right|_{\bar{\phi}=\varphi,\, G= {\cal G}}= 0 \,\, ,\\
\left. \frac{\delta V[\bar{\phi},G]}{\delta G(k)} \,
\right|_{\bar{\phi}=\varphi,\, G= {\cal G}}= 0 \,\, . \label{stationG}
\eea
With eq.\ (\ref{14}), the latter can be written in the form
\be \label{111}
{\cal G}^{-1}(k) = D^{-1}(k;\varphi) + \Pi(k)\,\, ,
\ee
\end{mathletters}
where
\be
\label{selfenergy}
\Pi(k) \equiv 2 \left. \frac{\delta V_2 [\bar{\phi},G]}{\delta G(k)} 
\right|_{\bar{\phi}=\varphi,\, G= {\cal G}} 
\ee
is the self energy.
Equation (\ref{111}) is the aforementioned Schwinger--Dyson equation.
The thermodynamic pressure is then determined by
\be
p = - V [\varphi,{\cal G}]\,\, ,
\ee
which, in the absence of conserved charges, is (up to a sign) identical to
the free energy density.

\section{The $O(N)$ Model} 

Let us now turn to the discussion of the $O(N)$ model. Its 
Lagrangian is given by 
\be 
{\cal L} =\frac{1}{2} \left( \partial_{\mu} \underline{\phi} \cdot 
\partial^{\mu} \underline{\phi} -  m^2 \, \underline{\phi} \cdot
\underline{\phi} \right) -  
\frac{\lambda}{N} \left(\underline{\phi} \cdot \underline{\phi} \right)^2 
 + H\, \phi_1\,\, , 
\ee
where $\underline{\phi}$ is an $N$-component scalar field.
For $H=0$ and $m^2 >0$, the Lagrangian is invariant under $O(N)$ rotations of 
the fields. For $H=0$ and $m^2 <0$, this symmetry is spontaneously 
broken down to $O(N-1)$, with $N-1$ Goldstone bosons (the pions).
The phenomenological explicit symmetry breaking term, $H$, is introduced to 
yield the observed finite masses of the pions.  
Spontaneous symmetry breaking leads to a non-vanishing 
vacuum expectation value for $\underline{\phi}$: 
\be 
\left| \langle\underline{\phi} \rangle \right| = \phi > 0\,\, .
\ee
($\phi$ assumes the role of $\bar{\phi}$ in section \ref{seceffpot}.)
At tree level,
\be \label{phi2}
 \phi \equiv f_\pi =  \sqrt{ - \frac{N\, m^{2}}{4\, \lambda}} \,
        \frac{2}{\sqrt{3}} \, \cos \frac{\theta}{3} \;\;\;,\;\;\;\; 
        \theta = \arccos \left[ \frac{H N}{8 \lambda} \left( 
        -\frac{12 \lambda}{N m^2} \right)^{3/2} \right] \,\, .
\ee
For $H=0$, $\cos(\theta/3) = \sqrt{3} / 2$. 
The inverse tree-level sigma and pion propagators are given by 
\begin{mathletters} 
\bea  \label{Dsigma}
D^{-1}_{\sigma}(k;\phi) &=& -k^{2} + m^{2} + 
        \frac{12\, \lambda}{N}\, \phi^{2} \,\, , \\ 
D^{-1}_{\pi}(k;\phi) &=& -k^{2} + m^2 + 
        \frac{4\, \lambda}{N}\, \phi^{2} \,\, . \label{Dpi}
\eea
\end{mathletters} 
This leads to the zero-temperature tree-level masses
\begin{mathletters} \label{vacuummasses}
\bea 
m^2_{\sigma} & = &m^{2} + \frac{12\, \lambda f^{2}_{\pi}}{N} \,\, ,\\ 
m^2_{\pi} & = & m^{2} + \frac{4 \, \lambda f^{2}_{\pi}}{N}
\eea
\end{mathletters}
for the sigma meson and the pion.
At tree level, the parameters of the Lagrangian are fixed 
such that these masses agree
with the observed values of $m_{\sigma}=$ 600 MeV and 
$m_{\pi} =$ 139 MeV.  Then, the coupling constant is 
\be \label{couplingconstant}
\lambda = \frac{N\, (m_{\sigma}^{2}-m_{\pi}^{2})}{8\, f^{2}_{\pi}} \,\, ,
\ee 
where $f_{\pi}=$ 93 MeV is the pion decay constant and 
\be 
m^{2} = -\frac{m_{\sigma}^{2}-3\, m_{\pi}^{2}}{2} \,\, .
\ee
The explicit symmetry breaking term is $H = m^{2}_{\pi} f_{\pi}$.
These tree-level results may change upon renormalization. 

\begin{figure}
\centering
\mbox{\epsfig{file=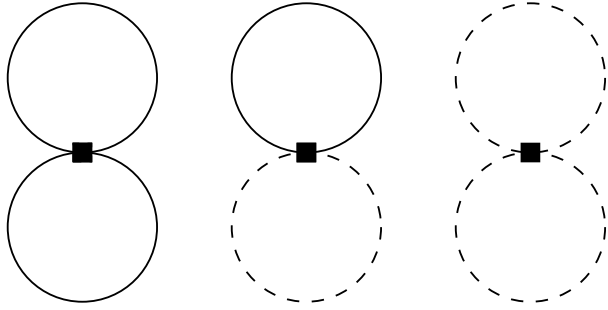,height=8cm,angle=270}}
\vspace*{0.5cm}
\caption{The Hartree contributions to the CJT effective potential. Full lines 
correspond to $G_\sigma$, while dashed lines correspond to $G_\pi$. The 
four-particle
vertex $\sim \lambda$ is represented by a full square.}
\label{fig:dbubble}
\end{figure}

The CJT effective potential for the $O(N)$ model is obtained 
from eq.\ (\ref{14}) as
\bea
V(\phi,G_{\sigma},G_{\pi})
 & = & \half m^{2} \phi^{2} + 
        \frac{ \lambda}{N} \phi^{4} - H \phi \nonumber \\
 & + & \half \int_k \left[ \ln G^{-1}_{\sigma}(k) + 
        D_{\sigma}^{-1}(k;\phi)\, G_{\sigma}(k)-1 \right] \nonumber \\
 & + &  \frac{N-1}{2} \int_k \, \left[
        \ln G^{-1}_{\pi}(k) + D_{\pi}^{-1}(k;\phi)\, G_{\pi}(k)-
        1\right] \nonumber \\
 & + &  V_{2}(\phi,G_{\sigma},G_{\pi})\,\, ,
\eea
where $V_{2}(\phi,G_{\sigma},G_{\pi})$ denotes the contribution from 
two-particle irreducible diagrams.  In the following we include only the 
two-loop diagrams shown in Fig.\ \ref{fig:dbubble} in $V_2$. These diagrams 
have no explicit $\phi$ dependence. Then, using eq.\ (\ref{selfenergy})
only tadpole diagrams (with resummed propagators) contribute to the
self energies. As explained in the introduction, this corresponds to the 
Hartree approximation. The Schwinger--Dyson 
equations for the full propagators contain
no momentum dependence. Thus, these equations are simply 
gap equations for the masses of the sigma meson and pion.

\begin{figure}
\centering
\mbox{\epsfig{file=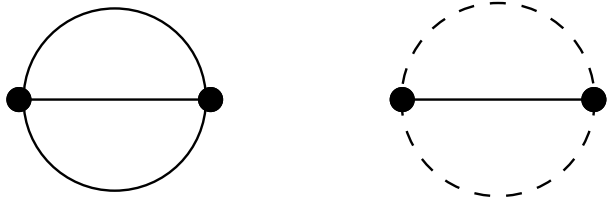,height=8cm,angle=270}}
\vspace*{0.5cm}
\caption{The neglected two-loop diagrams for the CJT effective potential. The
three-particle vertex $\sim \lambda \phi$ is represented by a full circle.}
\label{fig:sunset}
\end{figure}

On the two-loop level there exist, however, two more diagrams, cf.\
Fig.\ \ref{fig:sunset}, which will not be taken in our analysis. 
They depend explicitly on $\phi$
and introduce an additional momentum dependence in the Schwinger-Dyson 
equations, which makes their solution more complicated.
However, in the large-$N$ limit these terms are {\em a priori\/} absent,
because they are of order $1/N$.

In the Hartree approximation,
\bea
V_{2}(\phi,G_{\sigma},G_{\pi}) &=& 3\, \frac{\lambda}{N} 
        \left[ \int_k \, G_{\sigma}(k) \right]^{2} + 
        (N+1)(N-1)\, \frac{\lambda}{N} \left[
        \int_k \, G_{\pi}(k)\right]^{2} 
\nonumber \\
   & + & 2\,(N-1) \frac{\lambda}{N}
        \left[\int_k\, G_{\sigma}(k) \right] \left[\int_k \,
        G_{\pi}(k) \right] \,\, .
\eea
The coefficients in this equation are chosen such that, when computing the 
self energies from eq.\ (\ref{selfenergy}) and replacing the dressed
propagators by the tree-level propagators, one obtains the standard 
results for the perturbative one-loop self energies \cite{Peskin}.  
\newpage
\vspace*{-2cm}

\section{The Stationarity Conditions for the Effective Potential}

The stationarity conditions (\ref{stationphi}), (\ref{stationG}) read
\begin{mathletters}
\bea
0 & = &   m^2 \varphi + \frac{4 \lambda}{N}\, \varphi^3 - H
 +  \frac{4\lambda}{N}\, \varphi \int_q\,
\left[ 3 \,{\cal G}_\sigma(q) + (N-1) \, {\cal G}_\pi(q)\right] \,\, ,  \\
{\cal G}_\sigma^{-1}(k) & = & D_{\sigma}^{-1}(k;\varphi) + \frac{4\lambda}{N}
\int_q\, \left[ 3\, {\cal G}_\sigma(q) +(N-1)\, 
{\cal G}_\pi(q) \right] 
\,\, ,\\
{\cal G}_\pi^{-1}(k) & = & D_{\pi}^{-1}(k;\varphi) + \frac{4\lambda}{N}
\int_q\, \left[ {\cal G}_\sigma(q) +
(N+1)\, {\cal G}_\pi(q) \right] \,\, .
\eea
\end{mathletters}
The integrals on the right-hand side of the last two equations 
correspond to the sigma meson and pion self energies. According to eq.\ 
(\ref{selfenergy}), they originate from
the diagrams of Fig.\ \ref{fig:dbubble} via cutting one of 
the two loops in these diagrams. 
As one observes, these terms are independent of the
momentum $k^\mu$ appearing the propagator. 
The only $k$ dependence on the right-hand side enters through
$D_\sigma^{-1}$ and $D_\pi^{-1}$, cf. eqs.\ (\ref{Dsigma}) and (\ref{Dpi}).
Therefore, one is allowed to make the following {\em ansatz} for the full 
propagators:
\be \label{ansatz}
{\cal G}_{\sigma,\pi}(k) = \frac{1}{-k^{2} + M_{\sigma,\pi}^{2}} \,\, ,
\ee
where now $M_{\sigma}$ and $M_\pi$ are the masses dressed by interaction 
contributions from
the diagrams of Fig.\ \ref{fig:dbubble}.
Note that the diagrams in Fig.\ \ref{fig:sunset} have an
explicit dependence on the external momentum; including them would 
invalidate the ansatz (\ref{ansatz}).

The dressed sigma and pion masses are then determined by the following gap 
equations
\begin{mathletters}
\bea \label{Msigma}
M_{\sigma}^{2} & = & m^2 + \frac{4\lambda}{N} \left[
3\,\varphi^{2}  + 3\, Q(M_{\sigma},T)+ (N-1)\, Q(M_{\pi},T) \right] \,\, , 
\\
M_{\pi}^{2} & = & m^2 + \frac{4\lambda}{N} \left[ \varphi^{2} 
+ Q(M_{\sigma},T)+ (N+1)\, Q(M_{\pi},T) \right] \,\, . \label{Mpi}
\eea
\end{mathletters}
Here we introduced the function
\be \label{Q}
Q(M,T) \equiv  \int_k \, \frac{1}{-k^{2}+M^{2}}  
 =  \int \frac{d^3{\bf k}}{(2 \pi)^3}\,
    \frac{1}{\epsilon_{\bf k}(M)} \, \left\{ 
    \frac{1}{\exp[\epsilon_{\bf k}(M)/T]-1} + \half \right\}\,\, ,   
\ee
where $\epsilon_{\bf k}(M) \equiv \sqrt{{\bf k}^2 +M^2}$.
The last term in the integral is divergent and requires renormalization.
This will be discussed in section \ref{renorm}. The
standard practice, however, is to ignore this term, claiming it is independent
of temperature. This is wrong, because $\epsilon_{\bf k}(M)$ 
depends on $T$ through the gap equation for $M$.
As is shown below, the correct renormalization procedure 
changes the results.

Finally, $\varphi$ is determined by
\begin{mathletters}
\be  \label{phi3}
H = \varphi \left\{ m^2  + \frac{4 \lambda}{N} \left[ \varphi^2 
 + 3\, Q(M_\sigma,T) + (N-1) \, Q(M_\pi,T) \right] \right\}\,\, .
\ee
Using eq.\ (\ref{Msigma}) this can be written in the compact form
\be \label{phi4}
H = \varphi \left[M_\sigma^2 - \frac{8 \lambda}{N} \, \varphi^2 \right]\,\, .
\ee
\end{mathletters}
Note that this equation does not require explicit renormalization, and 
therefore is valid independent of the renormalization scheme.
Equations (\ref{Msigma}), (\ref{Mpi}), and (\ref{phi4}) are the 
stationarity conditions in the Hartree approximation.
In the case where chiral symmetry is not explicitly broken,
$H=m_\pi=0$, they imply the following: 
\begin{enumerate}
\item $\varphi \neq 0$. 
This is the phase where chiral symmetry is spontaneously broken. From 
eq.\ (\ref{phi4}) follows
\be \label{Msigma2}
M_\sigma = \sqrt{ \frac{8 \lambda}{N}} \, \varphi \,\, .
\ee
On the other hand, eqs.\ (\ref{Mpi}) and (\ref{phi3}) can be combined to
give
\be \label{Mpi2}
M_\pi^2 =  \frac{8 \lambda}{N} \left[ Q(M_\pi,T) - Q(M_\sigma,T) \right] \,\, .
\ee
This implies that Goldstone's theorem cannot be satisfied in the Hartree
approximation at all temperatures: $M_\sigma \neq 0$ on account 
of (\ref{Msigma2}), therefore $M_\pi = 0$ is not a solution of
(\ref{Mpi2}). (Note, however, that after proper renormalization 
of the function $Q(M,T)$, 
$M_\pi$ can be chosen to be zero at one particular temperature, 
for instance $T=0$, but then will be nonzero for other values of $T$.)

\item $\varphi = 0$. In this phase chiral symmetry is restored and 
eqs.\ (\ref{Msigma}) and (\ref{Mpi}) can be combined to
\be \label{51}
M_\sigma^2 - M_\pi^2 = \frac{8 \lambda}{N} \left[ Q(M_\sigma,T) -
Q(M_\pi,T) \right] \,\, ,
\ee
which has the solution $M_\sigma = M_\pi$; the masses become degenerate.
\end{enumerate}

Let us now turn to the discussion of the large-$N$ approximation which
is in fact the $N \gg 1$ limit of the Hartree approximation.
To derive the large-$N$ limit from the previous results, one simply neglects
all contributions of order $1/N$.
Note, however, that $\varphi^2 \sim N$, cf.\ eq.\ (\ref{phi2}). 
Therefore, in the large-$N$ limit, the stationarity conditions read
\begin{mathletters} \label{largeN}
\bea \label{52a}
M_{\sigma}^{2} &=& m^2 + \frac{4\lambda}{N} \left[
3\,\varphi^{2} + N\, Q(M_{\pi},T) \right]\,\, , \\
M_{\pi}^{2} &=& m^2 + \frac{4\lambda}{N} \left[\varphi^{2} +
        N\, Q(M_{\pi},T) \right] \,\, ,  \label{52b} \\
H & = & \varphi \left\{ m^2 + \frac{4 \lambda}{N} \left[\varphi^2 
+ N\, Q(M_\pi,T) \right] \right\} \, \, . \label{52c}
\eea
This leads to
\be \label{otherconsequ}
M_\sigma^2 = M_\pi^2 + \frac{8\lambda}{N}\, \varphi^2 \,\, ,
\ee
and
\be \label{52d}
M_\pi^2 \, \varphi = H \,\, .
\ee
\end{mathletters}
These two equations are valid independent of the renormalization scheme.
The latter equation implies the following in the case that chiral 
symmetry is not explicitly broken, $H=m_\pi=0$:
\begin{enumerate}
\item $\varphi \neq 0$. In this phase of spontaneously broken chiral symmetry
$M_\pi$ {\em has to vanish}, i.e., Goldstone's theorem is respected in the 
large-$N$ limit. $M_\sigma$ obeys the same relation (\ref{Msigma2}) as in
the Hartree approximation.
\item $\varphi = 0$. In this phase of restored chiral symmetry 
we again have $M_\sigma = M_\pi$, as in the Hartree case.
\end{enumerate}

\section{The Renormalized Gap Equations} \label{renorm}

As mentioned above, the last term in the integrand in (\ref{Q})
is divergent and requires renormalization.
In the following we discuss renormalization with a three-dimensional
momentum cut-off (CO) and via the counter-term (CT)
renormalization scheme.

In the literature one often encounters the argument that
this divergent term does not depend on temperature and can therefore 
either be absorbed in the definition of the renormalized vacuum mass 
in the CO scheme, or it is completely cancelled by a counter term in 
the CT scheme. This argument is correct to one-loop order in perturbation
theory, since then the mass $M$ in this term is simply the
bare mass and independent of temperature. 
However, in a self-consistent approximation scheme, like the
Hartree approximation, this argument is incorrect, since 
the mass $M$ is the {\em resummed\/} mass, which becomes a function of the 
temperature through the self-consistent solution of the gap equation.
Therefore, removing the divergence in either the CO or CT scheme
may leave a finite, temperature-dependent contribution. Another way
of stating this fact is that, as mentioned in the introduction, the
renormalization constants may have to be chosen such that they
depend on properties of the medium, like the temperature.

\subsection{CO scheme} 

The simplest way to regularize the divergent integral is to introduce
a three-dimensional ultraviolet momentum cutoff, $\Lambda$. 
Computing the divergent integral then 
proceeds as follows, 
\be
Q_\Lambda(M) \equiv 
\int^\Lambda \frac{d^{3}{\bf k}}{(2 \pi)^{3}} \frac{1}{2 \epsilon_{\bf k}(M)}
 =  \frac{1}{4 \pi^2} \int_0^\Lambda d k\, \frac{k^2}{\epsilon_{\bf k}(M)}
 =  \frac{1}{8 \pi^2} \left[ \Lambda \, \epsilon_\Lambda(M) - M^2
   \ln \frac{ \Lambda + \epsilon_\Lambda(M)}{M} \right] \,\, .
\ee
In the limit $\Lambda \rightarrow \infty$, this yields
\be \label{QM0}
Q(M,0) = \lim_{\Lambda \rightarrow \infty} Q_\Lambda(M) = 
\int \frac{d^{3}{\bf k}}{(2 \pi)^{3}} \frac{1}{2 \epsilon_{\bf k}(M)}
= I_1 - M^2 \, I_2 + \frac{M^2}{16 \pi^2} \ln \frac{M^2}{\mu^2} \,\, ,
\ee
where we have introduced a renormalization scale, $\mu$, and following 
\cite{Jackiw1} we have defined 
\begin{mathletters}
\bea 
I_{1} &\equiv& \lim_{\Lambda \rightarrow \infty} 
\frac{\Lambda^{2}}{8\pi^{2}}\,\, , \label{eq:I1}\\ 
I_{2} &\equiv& 
\lim_{\Lambda \rightarrow \infty} \frac{1}{16\pi^{2}}\, \ln
        \frac{4\, \Lambda^{2}}{\mu^{2}}\,\, .\label{eq:I2}  
\eea
\end{mathletters}
The renormalization is carried out by introducing new parameters 
\cite{Amelino}
\begin{mathletters}
\bea
\frac{m^{2}_{R}}{\lambda_{R}} & = & \frac{m^{2}}{\lambda} + 
        \frac{4(N+2)}{N}\, I_{1} \,\, , \\ 
\frac{1}{\lambda_{R}} & = & \frac{1}{\lambda} + \frac{4(N+2)}{N}\, I_{2} \,\, ,
\eea \label{mlren}
\end{mathletters}
where $m^{2}_{R}$ and $\lambda_{R}$ are the finite, renormalized mass and 
coupling constant. 

\subsubsection{Hartree approximation}

In the Hartree approximation, this leads to the following
renormalized gap equations for the sigma and pion masses:
\begin{mathletters}
\bea
M_\sigma^2 & = & m_R^2 + \frac{4 \lambda_R}{N} \, \frac{N+2}{N}
\left[ \varphi^2 + P(M_\sigma,T) + (N-1) \, P(M_\pi,T) \right]
 \nonumber \\
& - & \frac{2\, \lambda}{N\, \lambda_R} \left\{ M_\sigma^2 - m_R^2
- \frac{4 \lambda_R}{N}\,(N+2) \left[\varphi^2 + P(M_\sigma,T) 
\right] \right\}
\,\, , \label{MsigmaGAC} \\
M_\pi^2 & = & m_R^2 + \frac{4 \lambda_R}{N} \,\frac{N+2}{N} \left[
\varphi^2 + P(M_\sigma,T) + (N-1) \, P(M_\pi,T) \right]
\nonumber \\
& - & \frac{2\, \lambda}{N\, \lambda_R} \left\{ M_\pi^2 - m_R^2
- \frac{4 \lambda_R}{N}\,(N+2)\, P(M_\pi,T) \right\} \,\, , \label{MpiGAC}
\eea
\end{mathletters}
where the function $P(M,T)$ is defined as
\be \label{Q2}
P(M,T) = \frac{M^2}{16 \pi^2} \, \ln \frac{M^2}{\mu^2} + 
\int \frac{d^3{\bf k}}{(2 \pi)^3}\,   \frac{1}{\epsilon_{\bf k}(M)} \, 
    \frac{1}{\exp[\epsilon_{\bf k}(M)/T]-1} \,\, .
\ee
Equations (\ref{MsigmaGAC}) and (\ref{MpiGAC}) are 
equivalent to eqs.\ (13), (14) in \cite{Amelino} after the replacements
$\lambda \rightarrow \lambda /6\, , \, \phi^2 \rightarrow N\, \phi^2
\, , \, P(M,T) \rightarrow P_f[M]$. [Note that the terms $\sim M_\sigma^2
- M_\pi^2$ in eqs.\ (13), (14) of \cite{Amelino} can be eliminated by taking
the difference of eqs.\ (13) and (14).]

In the limit $\Lambda \rightarrow \infty$, $\lambda \rightarrow 0^-$ in order
to have a finite $\lambda_R$, and the (bare) theory becomes unstable
(see also Ref.\ \cite{Bardeen}).
Also, the (renormalized) masses obey $M_\sigma^2 = M_\pi^2$, cf.\ 
(\ref{MsigmaGAC}), (\ref{MpiGAC}), which is undesirable.
It would imply that chiral symmetry is 
unbroken, even when $\varphi \neq 0$. This problem was 
also addressed by the authors of \cite{Roh}.
On the other hand, for $0 < \lambda < \infty$, $\lambda_R \rightarrow 0^+$
in the limit $\Lambda \rightarrow 0$, indicating that the (renormalized)
theory becomes trivial \cite{Stevenson}.

Therefore, in the CO scheme the gap equations can only be meaningfully
studied for {\em finite\/} $\Lambda$. In this case, in the original gap 
equations (\ref{Msigma}), (\ref{Mpi}) we replace
\be \label{approx}
Q(M,T) \rightarrow Q_\Lambda (M) + Q_T (M)\,\, ,
\ee
where
\be
Q_T(M) \equiv Q(M,T)- Q(M,0) = 
\int \frac{d^3{\bf k}}{(2 \pi)^3}\,  \frac{1}{\epsilon_{\bf k}(M)} \, 
    \frac{1}{\exp[\epsilon_{\bf k}(M)/T]-1} \,\, .
\ee
The integral $Q_T(M)$ is UV-finite and does not require
the introduction of a momentum cut-off.
Consequently, the gap equations read
\begin{mathletters}
\bea \label{MsigmaCOH}
M_{\sigma}^{2} & = & m^2 + \frac{4\lambda}{N} \left\{
3\,\varphi^{2}  + 3\, \left[ Q_T(M_{\sigma}) +
Q_\Lambda(M_\sigma) \right] + (N-1)\, 
\left[ Q_T(M_{\pi}) + Q_\Lambda(M_\pi) \right] \right\} \,\, , 
\\
M_{\pi}^{2} & = & m^2 + \frac{4\lambda}{N} \left\{ \varphi^{2} 
+ \left[ Q_T(M_{\sigma}) +
Q_\Lambda(M_\sigma) \right] + (N+1)\, 
\left[ Q_T(M_{\pi}) + Q_\Lambda(M_\pi) \right] \right\} \,\, . \label{MpiCOH}
\eea
\end{mathletters}
The cut-off $\Lambda$ has to be determined from the values of $M_\sigma$
and $M_\pi$ at $T=0$:
\begin{mathletters}
\bea
m_\sigma^2 & = & m^2 + \frac{4 \lambda}{N} 
\left[ 3\, f_\pi^2 + 3\, Q_\Lambda(m_\sigma) + (N-1) \, Q_\Lambda
(m_\pi) \right]
\,\, , \label{MsigmaLR} \\
m_\pi^2 & = & m^2 + \frac{4 \lambda}{N} \left[
f_\pi^2 + Q_\Lambda(m_\sigma) + (N+1) \, Q_\Lambda(m_\pi) \right]
\,\, , \label{MpiLR} 
\eea
\end{mathletters}
where we have used $\varphi (T=0) = f_\pi$.
In the chiral limit ($H=m_\pi=0$), the difference of (\ref{MsigmaLR})
and (\ref{MpiLR}) reads
\be
m_\sigma^2 = \frac{8\lambda}{N} \left[ f_\pi^2 +Q_\Lambda(m_\sigma)
-Q_\Lambda(0) \right] \,\, .
\ee
However, from the stationarity condition (\ref{phi4}) we conclude
that $m_\sigma^2 = 8\, \lambda \, f_\pi^2/N$ in the chiral limit.
This immediately leads to
\be
Q_\Lambda(m_\sigma) = Q_\Lambda(0) \,\, ,
\ee
which for finite $m_\sigma$ can only be fulfilled if $\Lambda = 0$.
This, however, is exactly the case treated in \cite{Petro}, 
without renormalization.

In conclusion, the CO scheme fails to provide a consistent renormalization
of infinities in the phase of broken chiral symmetry in the Hartree
approximation when $H= m_\pi = 0$. Note that the same conclusion can be
reached with a four-dimensional momentum cut-off. This failure can
be traced to the fact that in the Hartree approximation 
diagrams of the type shown in Fig.\ \ref{fig:sunset} are not included
(cf.\ the perturbative renormalization of the linear sigma model 
\cite{Peskin}, see also the discussion in \cite{GB}).

Due to this failure, no results will be shown for
the Hartree approximation with CO renormalization.
However, we note that the case of explicitly broken symmetry,  $H \neq 0 , 
\, m_\pi>0$, is free of this problem. Then, the difference of eqs.\
(\ref{MsigmaLR}) and (\ref{MpiLR}) determines the coupling constant as
\be
\lambda \equiv \lambda(\Lambda) =  
\frac{N}{8} \, \frac{m_\sigma^2 -m_\pi^2}{f_\pi^2
+ Q_\Lambda(m_\sigma) - Q_\Lambda(m_\pi)} \,\, .
\ee
The mass parameter is given by
\be
m^2 = - \frac{m_\sigma^2 - 3\, m_\pi^2}{2} - \frac{4 \lambda}{N} \, 
(N+2)\, Q_\Lambda (m_\pi) \,\, .
\ee
$H$ is determined from (\ref{phi4}) to be
$H = f_\pi [m_\sigma^2 - 8 \lambda(\Lambda) f_\pi^2/N]$.

\subsubsection{Large-$N$ approximation}

In the large-$N$ limit, the
renormalized gap equation for the pion mass reads
\be
M_\pi^2 = m_R^2 + \frac{4 \lambda_R}{N} \left[ \varphi^2 + N\, P(M_\pi,T)
\right] \,\, ,
\ee
while $M_\sigma^2$ is still given by (\ref{otherconsequ}). This
again has the consequence that $M_\sigma = M_\pi$ in the limit
$\Lambda \rightarrow \infty$, i.e., $\lambda \rightarrow 0^-$, which
as discussed above is an unwanted feature. On the other hand, there is
no inconsistency in the large-$N$ approximation for finite $\Lambda$.
$\Lambda$ is a free parameter and the gap equations to be solved are 
(\ref{otherconsequ}) for the sigma mass and
\be \label{MpiCOLN}
M_{\pi}^{2}  =  m^2 + \frac{4\lambda}{N} \left\{ \varphi^{2} 
+ N\, \left[ Q_T(M_{\pi}) + Q_\Lambda(M_\pi) \right] \right\} \,\, 
\ee
for the pion mass.
The parameters are again determined from $M_\sigma(T=0) = m_\sigma$,
$M_\pi(T=0) = m_\pi$, and $\phi(T=0) = f_\pi$. From these 
conditions we derive that the coupling constant is still given by
its tree-level value, eq.\ (\ref{couplingconstant}), but $m^2$ is now 
determined from  
\be \label{65}
m^2 = - \frac{m_\sigma^2 - 3\, m_\pi^2}{2} - 4\, \lambda\, 
Q_\Lambda (m_\pi) \,\, .
\ee
$H$ retains its tree-level value on account of (\ref{52d}).

\subsection{CT scheme} 

In the CT scheme, counter terms are introduced to subtract the
UV divergences in $Q(M,T)$. Rewrite
\be \label{53}
\int \frac{d^{3}{\bf k}}{(2 \pi)^{3}} \frac{1}{2\, \epsilon_{\bf k}(M)}
\equiv \int \frac{d^{4}k}{(2 \pi)^{4}} \frac{1}{k^{2} + M^{2}}\,\, ,
\ee
where $k_0 \in {\bf R}$ with 
$k^2 = k_0^2 + {\bf k}^2$, $d^4k = d^3{\bf k}\, dk_0$.
To determine the counter terms, 
expand the integrand in a Taylor series around $M^2 = \mu^2$,
where $\mu$ is the renormalization scale.
\be
\frac{1}{k^2 + M^2} = \frac{1}{k^2 + \mu^2} 
\sum_{n=0}^\infty \left(\frac{\mu^2-M^2}{k^2+\mu^2} \right)^n 
\,\, .
\ee
The $d^4k$ integral over the $n=0$ term in this expansion is 
quadratically divergent, while the integral over the $n=1$ term diverges
logarithmically. The counter terms are chosen to remove these two terms,
such that the renormalized result for the divergent integral is
\be \label{55}
\int \frac{d^{4}k}{(2 \pi)^{4}} \left[ \frac{1}{k^{2} + M^{2}}
- \frac{1}{k^2 + \mu^2} - \frac{\mu^2-M^2}{(k^2 + \mu^2)^2} \right]
= \sum_{n=2}^{\infty} (\mu^2-M^2)^n \int \frac{d^{4}k}{(2 \pi)^{4}} \,
\frac{1}{(k^2+\mu^2)^{n+1}} \,\, .
\ee
Note that the second counter term depends on the temperature through $M$.
This fact represents the aforementioned possibility of having 
temperature-dependent counter terms in self-consistent approximation 
schemes, and was already discussed by the authors of \cite{GB}. They also 
pointed out that
this problem does not occur in less than three spatial dimensions.
This is obvious from eq.\ (\ref{55}), because then
the second counter term is finite, and thus not required.
In contrast, either in ordinary perturbation theory 
or in optimized perturbation theory \cite{Chiku} renormalization
at $T=0$ is sufficient to remove all divergences. 

The last integral in (\ref{55}) is finite and equal to
\be
\int \frac{d^{4}k}{(2 \pi)^{4}} \, \frac{1}{(k^2+\mu^2)^{n+1}}
= \frac{1}{(4 \pi)^2} \frac{\mu^{2(1-n)}}{n(n-1)} \,\, .
\ee
Expression (\ref{55}) can be rearranged to give the final result
\be \label{57}
\int \frac{d^{4}k}{(2 \pi)^{4}} \left[ \frac{1}{k^{2} + M^{2}}
- \frac{1}{k^2 + \mu^2} - \frac{\mu^2-M^2}{(k^2 + \mu^2)^2} \right]
= \frac{1}{(4 \pi)^2} \left[ M^2 \, \ln\frac{M^2}{\mu^2} - M^2 + \mu^2 \right]
\,\, .
\ee
To obtain the renormalized gap equations, simply replace $Q(M,T)$ as
given in (\ref{Q}) by
\be \label{71}
Q(M,T) = Q_T(M) + Q_\mu(M)\,\, ,
\ee
where
\be
Q_\mu(M) \equiv \frac{1}{(4 \pi)^2} \left[ M^2 \, \ln\frac{M^2}{\mu^2} - 
  M^2 + \mu^2 \right] \,\, .
\ee
The renormalization scale $\mu$ is chosen to give the correct values for
sigma and pion mass at $T=0$.
 
As an alternative to the above procedure, one
can also compute (\ref{53}) in dimensional regularization, i.e.,
in $d$ space-time dimensions, where the coupling constant 
$g$ is replaced by $g \tilde{\mu}^\epsilon$. Here, $\tilde{\mu}$
is the renormalization scale in dimensional regularization
and $\epsilon \equiv 4-d$.
In order to obtain (\ref{57}), one has to add a counter term
$M^2/(8 \pi^2 \epsilon) + \mu^2/(16 \pi^2)$.
Here, $\mu$ is the renormalization scale from the previous
treatment and related to $\tilde{\mu}$ by
$\mu^2 \equiv 4\pi e^{-\gamma} \tilde{\mu}^2$, where
$\gamma$ is the Euler-Mascheroni constant. Note again, that
the counter term depends implicitly on the temperature
through the resummed mass $M$.
In the Appendix, we furthermore show that the CO and CT schemes are
equivalent for unbroken $O(N)$ symmetry.

\subsubsection{Hartree approximation}

In the Hartree approximation, the 
gap equations read
\begin{mathletters}
\bea \label{MsigmaCTH}
M_{\sigma}^{2} & = & m^2 + \frac{4\lambda}{N} \left\{
3\,\varphi^{2}  + 3\, \left[ Q_T(M_{\sigma}) +
Q_\mu(M_\sigma) \right] + (N-1)\, 
\left[ Q_T(M_{\pi}) + Q_\mu(M_\pi) \right] \right\} \,\, , 
\\
M_{\pi}^{2} & = & m^2 + \frac{4\lambda}{N} \left\{ \varphi^{2} 
+ \left[ Q_T(M_{\sigma}) +
Q_\mu(M_\sigma) \right] + (N+1)\, 
\left[ Q_T(M_{\pi}) + Q_\mu(M_\pi) \right] \right\} \,\, . \label{MpiCTH}
\eea
\end{mathletters}
The renormalization scale $\mu$ is determined from
the vacuum values for the sigma and pion masses:
\begin{mathletters}
\bea
m_\sigma^2 & = & m^2 + \frac{4\lambda}{N} \left[ 3 f_\pi^2
+ 3\, Q_\mu(m_\sigma)+ (N-1)\, Q_\mu(m_\pi) \right]\,\, , \label{72a} \\
m_\pi^2 & = & m^2 + \frac{4\lambda}{N} \left[  f_\pi^2
+ Q_\mu(m_\sigma) + (N+1)\, Q_\mu(m_\pi)\right] \,\, .\label{72b}
\eea
\end{mathletters}
In the chiral limit, the difference of these two equations reads
\be
m_\sigma^2 = \frac{8\lambda}{N} \left[ f_\pi^2  + \frac{m_\sigma^2}{16 \pi^2}
\, \ln \frac{m_\sigma^2}{\mu^2 e} \right] \,\, .
\ee
However, in order to be consistent with the (generally valid)
equation (\ref{phi4}), there is only a single choice for the
renormalization scale, $\mu^2 \equiv m_\sigma^2 / e$. 
Then, the coupling constant is given by its classical value,
$\lambda = N\, m_\sigma^2/(8\, f_\pi^2)$, while
\be
m^2 = - \frac{m_\sigma^2}{2} - \frac{4 \lambda}{N}\, 
(N+2) \, \frac{\mu^2}{16 \pi^2} \,\, .
\ee

In the case that chiral symmetry is explicitly broken, the difference
of (\ref{72a}) and (\ref{72b}) yields the following equation for the
coupling constant:
\be
\lambda = \frac{N}{8} \, \frac{m_\sigma^2 - m_\pi^2}{f_\pi^2
+ \left[ m_\sigma^2 \ln(m_\sigma^2/\mu^2 e) - m_\pi^2 \ln(m_\pi^2/\mu^2 e)
\right]/16\pi^2 } \equiv \lambda(\mu) \,\, ,
\ee
i.e., $\lambda$ runs with the renormalization scale. However,
there is one value for the renormalization scale, where $\lambda$ retains
its tree-level (i.e.\ classical) value, 
\be
\mu^2 \equiv \mu_{\rm cl}^2 = 
\exp \left[ \frac{m_\sigma^2 \left( \ln m_\sigma^2 -1 \right)
- m_\pi^2 \left( \ln m_\pi^2 -1 \right)}{m_\sigma^2 - m_\pi^2} 
\right] \,\, .
\ee
The results for the Hartree case with explicitly broken symmetry
presented in the next section will exclusively employ this value of $\mu$.
The mass parameter is determined from
\be
m^2 = - \frac{m_\sigma^2 - 3\, m_\pi^2}{2} - \frac{4 \lambda}{N}\, 
(N+2) \, Q_\mu(m_\pi) \,\, .
\ee
$H$ can be obtained from (\ref{52d}) at $T=0$.

\subsubsection{Large-$N$ approximation}

In the large-$N$ limit, the gap equations to be solved are 
(\ref{otherconsequ}) for the sigma meson and
\be \label{MpiCTLN}
M_{\pi}^{2}  =  m^2 + \frac{4\lambda}{N} \left\{ \varphi^{2} 
+ N\, \left[ Q_T(M_{\pi}) + Q_\mu(M_\pi) \right] \right\} \,\, 
\ee
for the pion. In this case, $\mu$ is a free parameter, and cannot be
fixed by the vacuum values for the sigma and pion masses. 
$\lambda$ and $H$ are always given by their tree-level values.
The mass parameter is determined from
\be
m^2 = - \frac{m_\sigma^2 - 3\, m_\pi^2}{2} - 4\, \lambda\, 
\, Q_\mu(m_\pi) \,\, .
\ee

\section{Results} \label{results}

In this section, we discuss numerical solutions of the gap equations for
the meson masses and the stationarity condition on $\varphi$.
Three different cases are considered:
the large-$N$ approximation in (a) the CO scheme, (b) the CT scheme,
and (c) the Hartree approximation in the CT scheme. 
The Hartree approximation in the CO scheme will not be discussed, 
due to the problems exhibited in section \ref{renorm}. 
We focus separately on the cases $m_\pi = 0$ and $m_\pi >0$.

\subsection{\bf $m_{\pi}=0$} 

\begin{figure}
\hspace*{1cm}
\mbox{\epsfig{file=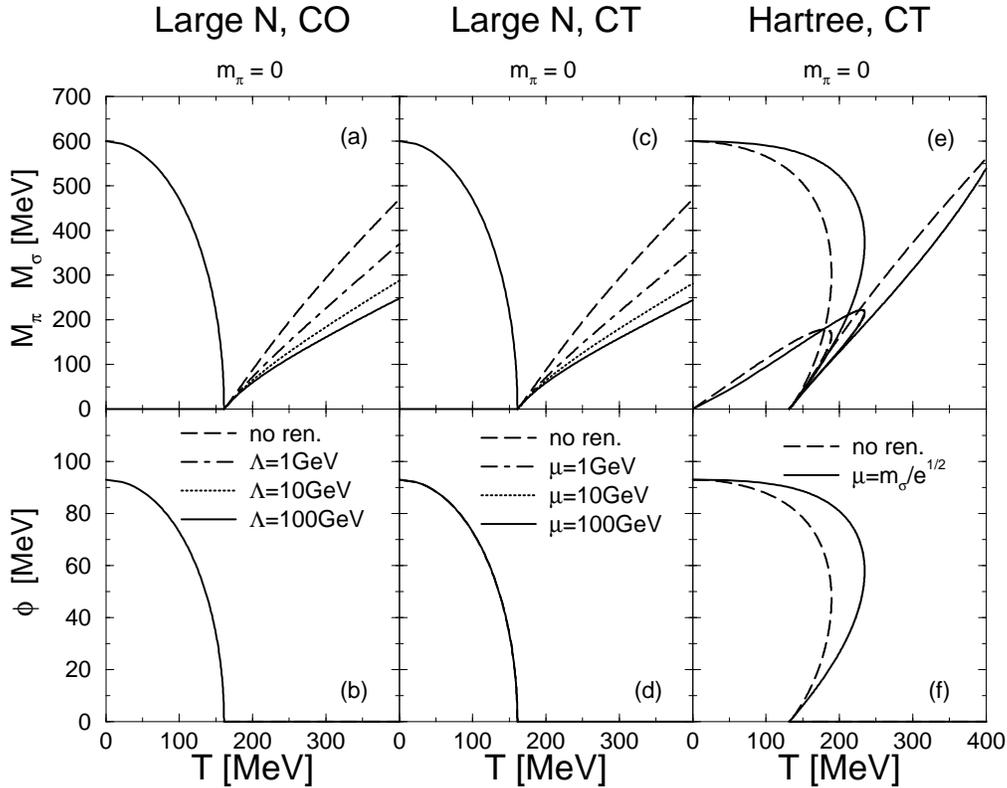,height=8cm,angle=270}}
\vspace*{-2cm}
\caption{The meson masses and $\varphi$ as a function of $T$ for 
$m_{\pi}=0$.\label{fig:massesmpi0}}
\end{figure}

Figures \ref{fig:massesmpi0} (a,c,e) show the meson masses and
(b,d,f) $\varphi$ as functions of temperature. Results for the
large-$N$ approximation with CO renormalization are shown in parts (a,b), 
and with CT renormalization in (c,d). Results for the 
Hartree approximation with
CT renormalization are shown in (e,f). For comparison, the dashed lines 
in each figure correspond to the unrenormalized results of \cite{Petro}.

In Figs.\ \ref{fig:massesmpi0} (a,b), in the phase of spontaneously
broken symmetry, there is no difference between the unrenormalized and
renormalized cases. To understand this, first remember that $M_\pi=0$,
cf.\ the discussion following eq.\ (\ref{52d}). Therefore, on account
of (\ref{otherconsequ}), $M_\sigma$ is simply given by $(8 \lambda
/N)^{1/2} \varphi$. In turn, $\varphi$ is determined by (\ref{52c}).
However, for $M_\pi=0$, this has the simple form
\be
0 = m^2 + \frac{4 \lambda}{N} \, \varphi^2 + 4 \lambda \left[ \frac{T^2}{12}
+ Q_\Lambda (0) \right] \,\, .
\ee
Using (\ref{65}) for $m_\pi=0$, this becomes
\be
0 = - \frac{m_\sigma^2}{2} + \frac{4 \lambda}{N} \, \varphi^2 
+ 4 \lambda \frac{T^2}{12}\,\, ,
\ee
which is the same condition as in the unrenormalized case (where $Q_\Lambda$
is absent). Since the coupling constant is given by 
its tree-level value (\ref{couplingconstant}), this immediately leads to 
the conclusion that the temperature for chiral symmetry restoration is
\be
T^* = \sqrt{3}\, f_\pi \,\, .
\ee
In the restored phase, $\varphi = 0$, sigma and pion masses are equal,
and given by eq.\ (\ref{52a}) or (\ref{52b}). These equations are cut-off
dependent, on account of (\ref{approx}). The mass is decreasing for
increasing $\Lambda$.

In Figs.\ \ref{fig:massesmpi0} (c,d), we show results for the
large-$N$ approximation in the CT scheme. In the broken phase, $\varphi
>0$, renormalization again does not affect the masses or $\varphi$.
In the phase of restored symmetry, $\varphi = 0$, the sigma and pion masses
are degenerate, but depend on the renormalization scale.
They decrease for increasing $\mu$. Note the similarity between
the masses in the CO and the CT scheme when choosing the same 
value for the cut-off $\Lambda$ and the renormalization scale $\mu$. 
Considering that both renormalization schemes are fundamentally different, 
this similarity is quite surprising.
Another important conclusion is that renormalization of the gap equations
does not destroy the second-order nature of the transition.

\vspace*{1.5cm}

The results in the Hartree approximation, eqs.\ (\ref{Msigma}) -- 
(\ref{phi4}), are
displayed in Figs.\ \ref{fig:massesmpi0} (e,f). As in the unrenormalized
case, we obtain a first order transition, with a transition temperature
that appears to be slightly higher than in the unrenormalized case. 
To determine this temperature, however, one would have to analyze
the shape of the effective potential, which is outside the scope of this
paper.
 
\subsection{\bf $m_{\pi} = 139$ MeV} 

\begin{figure}
\hspace*{1cm}
\mbox{\epsfig{file=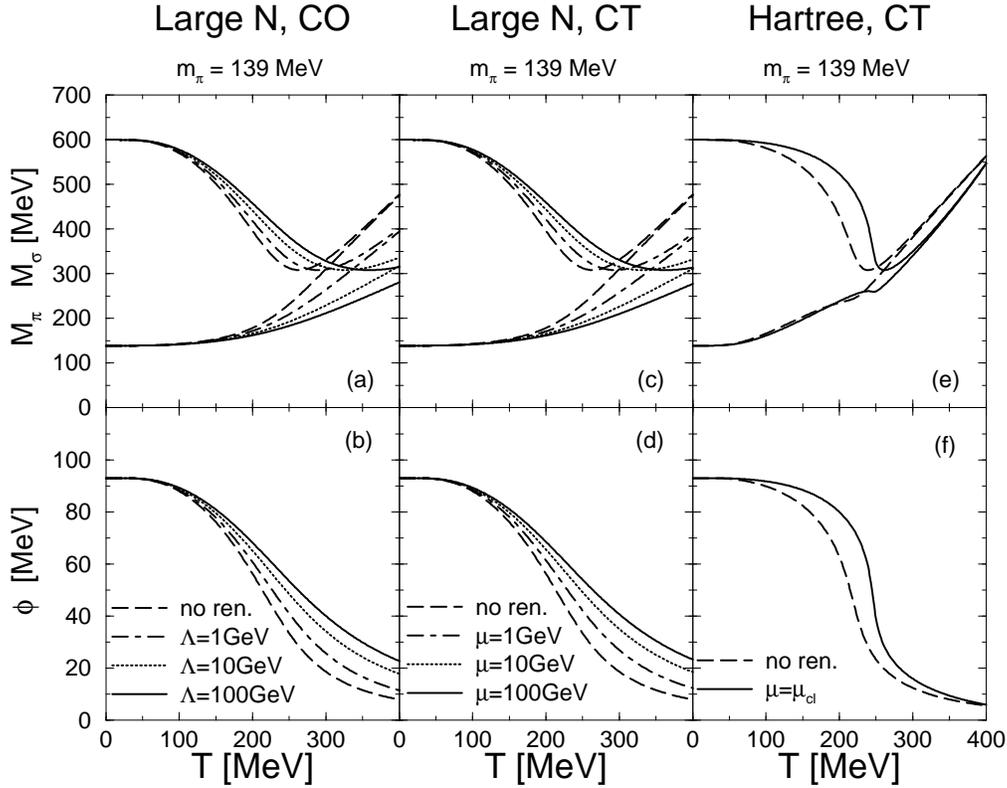,height=8cm,angle=270}}
\vspace*{-2cm}
\caption{The meson masses and $\varphi$ as a function of $T$ for 
$m_{\pi}=$139 MeV.\label{fig:masses2}}
\end{figure}

Fig.\ \ref{fig:masses2} (a,c,e) shows the temperature dependence of the
meson masses and Fig.\ \ref{fig:masses2} (b,d,f) the function $\varphi(T)$ 
in the case of explicit symmetry breaking. As in Fig.\
\ref{fig:massesmpi0}, large-$N$ results are shown in (a,b) for the 
CO scheme and in (c,d) for the CT scheme. Parts (e,f) show our results for the
Hartree approximation with CT renormalization.
As already observed in the chiral limit, there is a striking similarity
between the results in the CO and the CT schemes when choosing
$\Lambda = \mu$. Also, increasing the cut-off or the renormalization 
scale tends to increase the temperature
at which (approximate) symmetry restoration takes place.

Baym and Grinstein \cite{GB} noted that the additional terms originating
from renormalization have the effect that the gap equations do not have
a solution beyond a certain temperature (see also \cite{Chiku,Bardeen}). 
We found evidence for this
in the CT scheme at temperatures above 400 MeV. In the CO scheme with
a finite $\Lambda$, this phenomenon does not occur.

\vspace*{2cm}

\section{Conclusions and Outlook}

In this paper, we have studied 
the temperature dependence of sigma and pion masses
in the framework of the $O(N)$ model. The Cornwall--Jackiw--Tomboulis 
formalism was applied to derive gap equations for the masses
in the Hartree and large-$N$ approximations.
Renormalization of the gap equations was carried out within the cut-off and
counter-term renormalization schemes. 
In agreement with \cite{Amelino}, it was found that the cut-off scheme 
is flawed
when the cut-off $\Lambda \rightarrow \infty$. We therefore studied this
renormalization scheme for $\Lambda < \infty$. For the Hartree
approximation we found that, in the chiral
limit ($m_\pi = 0$), there is no finite value for the cut-off, which
is consistent with the set of
stationarity conditions for the effective potential;
$\Lambda = 0$ is the only possible choice. This problem
was not encountered in the large-$N$
approximation; here any choice for $\Lambda$ is possible.
In the counter-term renormalization scheme, the Hartree approximation can
be consistently renormalized, but in the chiral limit, the renormalization
scale is restricted to a unique value in order to achieve consistency
with the stationarity conditions for the effective potential. In the
large-$N$ limit, the renormalization scale can be chosen arbitrarily.
Changing the cut-off in the cut-off scheme or the renormalization 
scale in the counter-term scheme changes the meson masses at a given 
temperature.
The reason is that, in the self-consistent approximation schemes considered 
here, the renormalization constants (or counter terms, respectively) may 
depend implicitly on temperature. This does not occur when
renormalizing ordinary perturbation theory.

Our results can be compared to those of Roh and Matsui \cite{Roh} and
Chiku and Hatsuda \cite{Chiku}. The authors of \cite{Roh}
computed the sigma and pion masses from the second derivative of an
effective potential which was determined via the standard loop 
expansion approach.
Being aware that this approach fails for theories with spontaneously broken
symmetry, they corrected the resulting expressions to
obtain gap equations which look similar to
the ones in the Hartree approximation. (They are identical to
Baym and Grinstein's modified Hartree approximation \cite{GB}). 
The stationarity condition for
$\varphi$, however, was taken to be the same as in the large-$N$ approximation.
Thus, their solutions respect Goldstone's theorem in the phase of spontaneously
broken symmetry, similar to the large-$N$ approximation discussed here,
while the transition in their model is first order (in the chiral limit),
like in the Hartree approximation. 

The authors of \cite{Chiku} employ optimized perturbation theory
to compute the sigma and pion masses. This approach has the advantage
that renormalization is straightforward.
The results are similar to those of \cite{Roh}.

Recent dilepton experiments at CERN-SPS energies \cite{dileptons} 
have generated 
interest in medium modifications of meson properties such as their 
mass and decay width. In general, the meson mass (squared) is given 
by the inverse propagator at $k = 0$, 
$M^2 \equiv {\cal G}^{-1}(0)$.
The decay width of a particle with energy $\omega$ at rest,
${\bf k}=0$, is given by 
$\gamma(\omega) \equiv - {\rm Im} \Pi(\omega,{\bf 0})/\omega$ 
\cite{Weldon}, where $\Pi(\omega,{\bf k})$ is the self energy. 
In the CJT formalism, ${\cal G}^{-1}(k) = D^{-1}(k;\varphi) + \Pi(k)$, 
cf.\ eq.\ (\ref{111}).
In the Hartree or large-$N$ approximation studied here, the 
self energies do not acquire an imaginary part, because they are simply
constants, and thus {\em only\/} shift the mass of the particles.
Therefore, in these approximations, the particles are true 
quasi-particles with vanishing decay width.
This would change if we included the diagrams of
Fig.\ \ref{fig:sunset} in the effective potential, because, as is
well-known \cite{Weldon}, 
the imaginary part of these diagrams corresponds to decay and 
scattering processes.

To include these diagrams in the above treatment, however,
is prohibitively difficult, because then the simple momentum dependence
of the propagators ${\cal G}_{\sigma,\pi}(k)$ 
in eq.\ (\ref{ansatz}) changes, since the self energies 
become explicitly momentum dependent. Then, instead of simple gap equations
for the meson masses, the stationarity condition (\ref{stationG})
becomes an (infinite) set of coupled integral equations
for the propagators ${\cal G}_{\sigma,\pi}(k)$.

\begin{figure}
\hspace*{1.5cm}
\mbox{\epsfig{figure=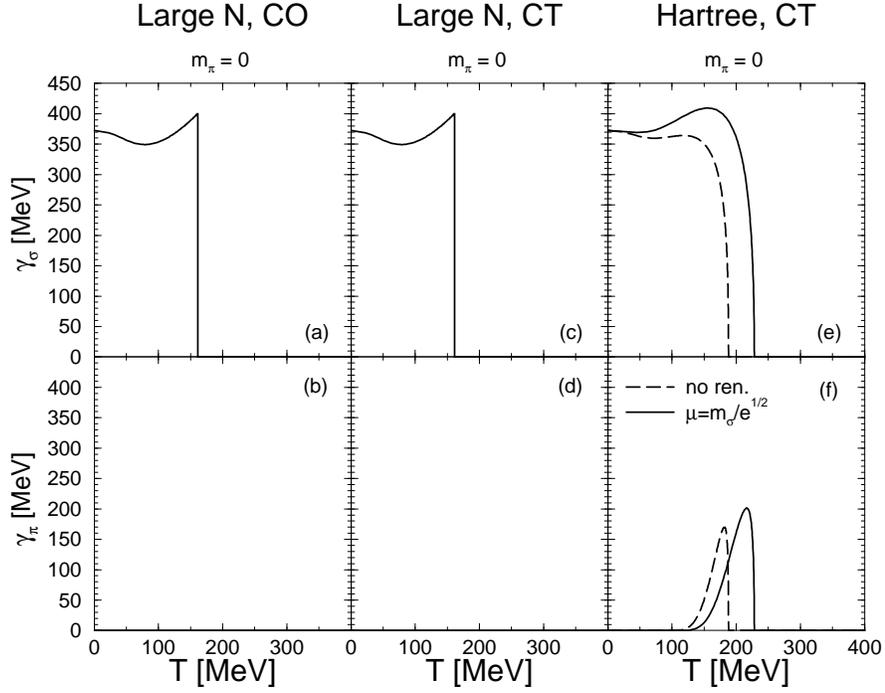,height=7cm,angle=270}}
\vspace*{-2cm}
\caption{The sigma and pion decay widths as a function of $T$ for 
$m_{\pi}=$0.\label{fig:mpi0eta}}
\end{figure}

\begin{figure}
\hspace*{1.5cm}
\mbox{\epsfig{figure=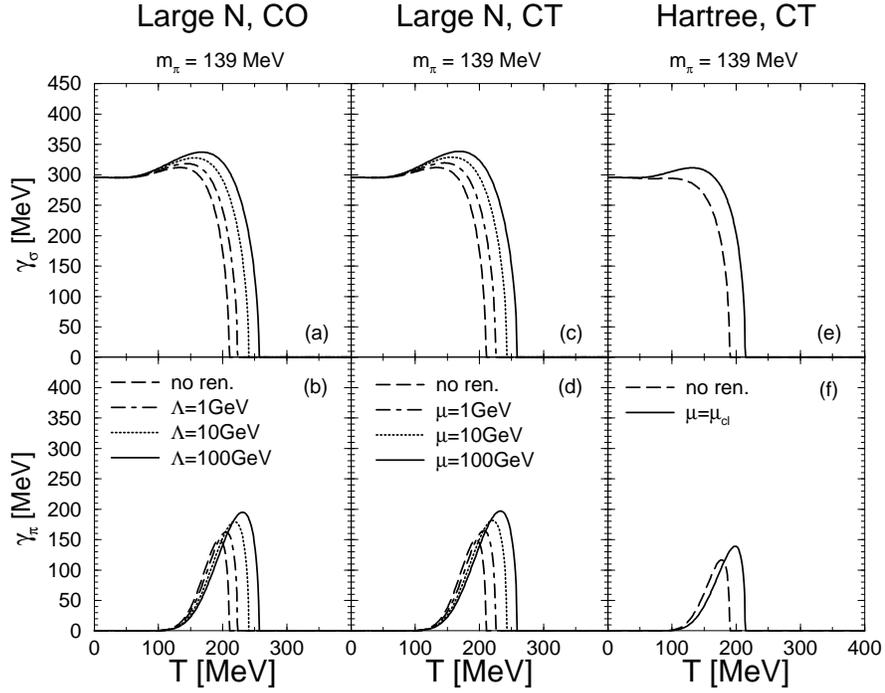,height=7cm,angle=270}}
\vspace*{-2cm}
\caption{The sigma and pion decay widths as a function of $T$ for 
$m_{\pi}=$139 MeV.\label{fig:mpi139eta}}
\end{figure}

Therefore, as a first approximation, we compute the decay widths
from the self energies
corresponding to these diagrams, but with internal lines  given by 
the Hartree or large-$N$ propagators (\ref{ansatz}).
This is equivalent to computing the decay width to one-loop order in
perturbation theory, but taking the medium-modified masses of the particles 
computed above instead of the vacuum masses.
The on-shell decay width of $\sigma$ and $\pi$ mesons at rest is then
given by the following expressions \cite{Dirk1}, valid for
$2\, M_\pi \leq M_\sigma$:
\begin{mathletters}
\bea
\gamma_{\sigma} & = & \left( \frac{4 \lambda \phi}{N} \right)^{2} 
        \frac{N-1}{16 \pi \, M_{\sigma}} \sqrt{1-\frac{4 M_{\pi}^{2}}
        {M_{\sigma}^{2}}} \, 
        {\rm coth} \, \frac{M_{\sigma}}{4T} \,\, , \\
\gamma_{\pi} & = & \left( \frac{4 \lambda \phi}{N} \right)^{2} 
        \frac{M_{\sigma}^{2}}{8 \pi \, M_{\pi}^{3}}  \,
        \sqrt{1-\frac{4 M_{\pi}^{2}}{M_{\sigma}^{2}}} \,
        \frac{1-\exp[-M_{\pi}/T]}{1-\exp[-M_{\sigma}^{2}/2m_{\pi}T]}\,
        \frac{1}{\exp[(M_{\sigma}^{2}-2M_{\pi}^{2})/2M_{\pi}T]-1} \,\, .
\eea
\end{mathletters}
These quantities are shown in Fig.\ \ref{fig:mpi0eta} for
$m_\pi = 0$, and in Fig.\ \ref{fig:mpi139eta} for $m_\pi = 139$ MeV,
for the cases discussed in Figs.\ \ref{fig:massesmpi0} and
\ref{fig:masses2}. For $m_\pi = 0$ and in the large-$N$ approximation, 
pions are true Goldstone bosons, and therefore their decay width
vanishes below the 
temperature corresponding to chiral symmetry restoration, see
Figs.\ \ref{fig:mpi0eta} (b,d). 
This is different in the Hartree approximation, where Goldstone's theorem
is violated, cf.\ Fig.\ \ref{fig:mpi0eta} (f), and when chiral
symmetry is explicitly broken, Figs.\ \ref{fig:mpi139eta} (b,d,f). 
The reason is that, because pions have a finite mass, they can acquire a 
finite decay width
on account of the absorption processes $\pi \sigma \rightarrow \pi$ and
$\pi \pi \rightarrow \sigma$. For massless particles, these processes
are kinematically forbidden.

Sigma mesons, however, can always decay into two pions, and therefore 
acquire a large decay width, cf.\ Figs.\ \ref{fig:mpi0eta} and
\ref{fig:mpi139eta} (a,c,e).
All decay widths vanish above the temperature where $M_\sigma$ becomes
smaller than $2\, M_\pi$. This, however, is an artefact of the one-loop
approximation. In two-loop order, the scattering processes
$\sigma \sigma \rightarrow\sigma \sigma,\,
\sigma \sigma \rightarrow \pi \pi,\, \sigma \pi \rightarrow \sigma \pi$,
and $\pi \pi \rightarrow \pi \pi$ lead to a finite decay width for all
particles even above this threshold.

The decay widths and masses computed here are relevant for the formation of
disoriented chiral condensates \cite{Dirk1}, since they enter 
the evolution equations of the long-wavelength modes.
This will be the subject of a subsequent investigation \cite{JTLDHR}.

\begin{center} 
{\bf Acknowledgements} 
\end{center}

We thank T.\ Appelquist, J.\ Berges, 
S.\ Gavin, M.\ Gyulassy, T.\ Hatsuda,
Y.\ Kluger, J.\ Knoll, L.\ McLerran,
E.\ Mottola, B.\ M\"uller, and R.\ Pisarski
for valuable discussions.
D.H.R.\ thanks Columbia University's Nuclear Theory group for continuing
access to their computational facilities.
J.T.L.\ is supported by the Director, Office of Energy
Research, Division of Nuclear Physics of the Office of 
High Energy and Nuclear Physics of the U.S.\ Department of 
Energy under Contract No.\ DE-FG-02-91ER-40609.

\newpage

\vspace*{1cm}

\appendix

\section{Scheme Equivalence for Unbroken $O(N)$ Symmetry}

In this Appendix, we show that the CO and the CT schemes are 
equivalent when the $O(N)$ symmetry is not broken ($\varphi = 0$).
In this case, the gap equations become degenerate,
\be \label{unrenormMsymm}
M^2 = m^2 + \frac{4 \lambda}{N}\, (N+2) \, Q(M,T) \,\, .
\ee
(We only discuss the Hartree case here, for the large-$N$ approximation, 
simply replace $N+2$ by $N$.)
In the CO scheme, using eqs.\ (\ref{QM0}) and (\ref{mlren}), this becomes
\be \label{Msymmetric}
M^2 = m_R^2 + \frac{4 \lambda_R}{N} (N+2) \left[ Q_{T}(M)
+ \frac{M^2}{16 \pi^2} \, \ln \frac{M^2}{\mu^2} \right] \,\, .
\ee
The renormalization scale $\mu^2$ can be determined from the
$T=0$ limit of this equation. For unbroken $O(N)$ symmetry,
$M(T=0) = m_R$, which then yields $\mu = m_R$.

On the other hand, in the CT scheme, we have
\be \label{MsymmCT}
M^2 = m_R^2 + \frac{4 \lambda_R}{N} (N+2) \left[ Q_{T}(M)
+ \frac{M^2}{16 \pi^2} \, \ln \frac{M^2}{\mu^2} 
- \frac{M^2-\mu^2}{16 \pi^2} \right] \,\, .
\ee
Here, we made the finiteness of 
$m$ and $\lambda$ explicit by replacing them with $m_R$ and $\lambda_R$. 
Again, the condition $M(T=0) = m_R$ yields $\mu = m_R$.

The last term in (\ref{MsymmCT}) leads to an apparent difference between 
the two schemes.
However, shifting the coupling constant by a finite amount,
\be
\frac{1}{\lambda_R} \rightarrow \frac{1}{\lambda_R} - 
\frac{4 (N+2)}{16 \pi^2 N}\,\, ,
\ee
one obtains (\ref{Msymmetric}), which proves the equivalence of both schemes
after properly redefining the coupling constant. 
The same conclusion can be reached starting
from (\ref{unrenormMsymm}) and using instead of (\ref{mlren}) the
modified renormalization conditions
\begin{mathletters}
\bea
m_R^2 \left[ \frac{1}{\lambda_R} + \frac{4 (N+2)}{16 \pi^2 N} \right]
& = & \frac{m^2}{\lambda} + \frac{4 (N+2)}{N} \, I_1 \,\, ,\\
\frac{1}{\lambda_R} + \frac{4 (N+2)}{16 \pi^2 N}
& = & \frac{1}{\lambda} + \frac{4 (N+2)}{N} \, I_2 \,\, ,
\eea
\end{mathletters}
which then leads to (\ref{MsymmCT}).

\end{document}